# A small-area ecological approach for estimating vote changes and their determinants

Bruno Bracalente [a], Antonio Forcina [a] e Nicola Falocci [b]

[a] *Department of Economics of the University of Perugia (Italy)*
[b] *Policy Evaluation and Control Service of the Legislative Assembly of the Umbria Region (Italy)*

**Abstract**

Empirical analyses on the factors driving vote switching are rare, usually conducted at the national level without considering the parties of origin and destination, and often unreliable due to the severe inaccuracy of recall survey data. To overcome the problem of lack of adequate data and to incorporate the increasingly relevant role of local factors, we propose an ecological inference methodology to estimate the number of vote transitions within small homogeneous areas and to assess the relationships between these counts and local characteristics through multinomial logistic models. This approach allows for a disaggregate analysis of contextual factors behind vote switching, distinguishing between their different origins and destinations. We apply this methodology to the Italian region of Umbria, divided into 19 small areas. To explain the number of transitions toward the right-wing nationalist party that won the elections and towards increasing abstentionism, we focused on measures of geographical, economic, and cultural disadvantages of local communities. Among the main findings, the economic disadvantages mainly pushed previous abstainers and far-right Lega voters to change their choices in favor of the rising right-wing party, while transitions from the opposite political camp were mostly influenced by cultural factors such as a lack of social capital, negative attitude towards the EU, and political tradition.

## 1. Introduction

In recent years, many countries have witnessed significant voter shifts from one election to the next. These changes have strongly influenced electoral outcomes, mostly weakening traditional parties and benefiting new populist parties and abstention. The growing volatility in voting behavior highlights the need to investigate vote switching across elections, rather than simply analyzing choices in a single election. However, research on the factors driving individual-level change is scarce and hampered by problems of data accuracy. The main empirical challenge lies in how to reliably observe or estimate vote switching, which requires information about individuals' previous and current voting behavior. Political scientists typically rely on respondents' recollections in post-election national surveys. Unfortunately, for various reasons, these recollections suffer from severe inaccuracies—especially for unstable voters and abstainers. As a result, both electoral volatility and abstention tend to be significantly underestimated (see Dejaeghere and Dassonneville, 2017). More importantly, if the inaccuracy is not random but linked to the same factors that influence vote choice and switching, it can severely distort estimates of those factors' effects (Bernstein et al., 2001; Schoen, 2011). Moreover, survey-based individual data become even less reliable when analysis aims to disaggregate the origins and destinations of vote change, rather than just distinguishing between stable and unstable voters. Despite their relevance, empirical studies of vote switching remain rare.

From a different perspective, many political scientists have suggested shifting the focus from individuals to local communities, using small geographical areas as units of analysis. This approach is particularly appropriate when studying issues with highly unequal spatial impacts—such as market globalization and economic restructuring—which tend to elicit similar electoral responses among voters in the same area (Broz et al., 2021, among others). However, empirical research at the local level is problematic using national survey data, due to the very limited number of observations (if any) within each small area.

To address these issues, following authoritative electoral scholars, we argue that the scarcity of appropriate data necessitates the use of ecological methods (Johnston and Pattie, 2006). In this study, we propose estimating voter transitions in small homogeneous areas using ecological inference methods—the only feasible approach at this geographic scale—and analyzing their contextual determinants through appropriate models. Using official electoral data recorded at the polling station level inside small homogeneous areas, the risk of bias (ecological fallacy) that has often discouraged the use of ecological inference it is minimized. Treating small areas as units of analysis also allows the inclusion of covariates related to the collective issues affecting local communities.

This article presents the proposed methodological approach and its application to the Italian region of Umbria. Voter transitions were estimated using a modified version of the Brown and Payne (1986) model within 19 small homogeneous areas, followed by estimation of vote transition counts. We then modeled the relationships between key vote transitions and local contextual variables using multinomial logistic regression. The goal is to analyze how the geographic, economic, and cultural disadvantages of local communities influenced voter transitions between the 2018 and 2022 national elections—both in favor of the nationalist right-wing party that won and in terms of rising abstention.

The remainder of the paper is organized as follows:

Section 2 reviews the theoretical and empirical literature on vote switching at both the individual and local community level, discusses the issue of sample data accuracy, and outlines the proposed approach. Section 3 describes the criteria used to define small areas and the statistical methodologies employed to estimate vote transitions and model their relationships with covariates. Section 4 presents the empirical application and main results. Section 5 provides discussion and concluding remarks.

## 2. A review of literature on voting changes and the proposed approach

**2.1 The accuracy of survey data on voting changes**

In many countries, electoral volatility has dramatically increased, reaching unprecedented levels in recent years. Using individual-level data from the Comparative Study of Electoral Systems (CSES) surveys, Dejaeghere and Dassonneville (2017) estimated party switching in 33 elections across 19 countries between 2002 and 2011. They found that, on average, 30% of the nearly 30,000 respondents reported having changed their party preference, with substantial variation across elections and countries. Fieldhouse et al. (2020), analyzing the British Election Study (BES) panel data, reported that in the UK, party switching increased from about 13% in the 1960s to 43% in the 2015 elections. As for the other component of total volatility—shifts from voting to abstention—available data are partial and largely inaccurate. Dassonneville et al. (2015), using CSES data from 36 elections across 22 countries (2001–2011), found that only 5.4% of the nearly 35,000 respondents reported switching to abstention, a figure the authors themselves considered underestimated.

To detect previous vote behavior, the CSES uses recall questions, unlike the BES panel, which directly tracks respondents across time. Numerous studies have shown that recall methods underestimate both party switching and abstention. Waldahl and Aardal (2000), based on Norwegian Program Electoral Research panel surveys, compared party choices declared immediately after the previous elections with those recalled four years later. They found that about 25% of respondents gave inaccurate answers, rising to 40% among party switchers and 70% among prior non-voters. Dassonneville and Hooghe (2017), using election panel surveys from Belgium, the Netherlands, and Germany, showed that switching voters consistently had much higher rates of erroneous reporting compared to stable voters—up to 65.6% in Belgium, depending on the time span between elections. Empirical studies confirm that recall error is not random. It is biased by a tendency to reconcile past and present vote choices and by the social desirability of certain responses, resulting in apparent vote stability (Waldahl and Aardal, 2000; Dassonneville and Hooghe, 2017).

Turnout data also suffer from severe inaccuracy. McAllister and Quinlan (2022), in a study covering 184 elections in 55 countries using CSES data, found that reported turnout was overestimated by almost 15%, with this bias increasing over time. Importantly, turnout overreporting is also non-random. Bernstein et al. (2001), comparing reported and validated turnout data from the American National Election Study (ANES), showed that those most likely to misreport abstention were the most pressured to vote—highly educated, partisan, or religious respondents. This misreporting of turnout and switching parties distorts models of voting behavior, biasing the estimated effects of independent variables and leading researchers to miss or falsely confirm hypotheses (see also Adida et al., 2019). These vote-specific inaccuracies add to general issues in survey research, such as self-selection bias. Particularly in recent years, voters for populist and extremist parties—recently the far-right—resulted under-represented in the sample surveys and under-declared by the respondents, contributing to the underestimation of the party switching (Durand et al., 2015; Pavia et al., 2016; Bodet et al., 2025). In Italy, Russo (2015) found that the ITANES post-election survey (2008) significantly underestimated abstention, arguably due to non-voters being less likely to participate in such surveys.

**2.2 The determinants of vote changes**

The literature attributes rising voter mobility to two main causes: (1) gradual changes in electoral behavior reflecting a more fluid electorate, and (2) external shocks or rapid structural transformations in the economy that influence large segments of voters (Fieldhouse et al., 2020; Guriev and Papaioannu, 2020). The first cause is mainly associated with the decline of partisan identification and the phenomenon of "partisan dealignment" (Dalton, 1984; Dassonneville, 2012). The replacement of older generations with strong party ties by younger cohorts with weak or no political affiliation is considered a key driver. Moreover, newer forms of individualized political engagement—particularly among young people—are increasingly based on post-materialist values (e.g., environmentalism, gender equality), which have partly replaced traditional social cleavages (Inglehart, 2008). The rise in vote switching and abstention is, at least partly, a consequence of these long-term shifts.

*The individual level of analysis*

Only a few empirical studies have explored the drivers of vote switching and electoral volatility. Notably, Dassonneville and co-authors have produced several contributions. Among these, Dassonneville et al. (2015) offer a relevant methodological approach. They analyze the probability that a voter in the previous election chooses one of three options: loyalty, party change, or abstention. Using multinomial logistic regression models on CSES data (2007 and 2013) from 36 elections in 22 countries, they find that dissatisfaction with specific parties and the number of available options both affect vote switching, while generalized dissatisfaction with the political system is the main factor behind abstention.

Economic conditions also matter. Although most economic voting literature focuses on incumbent performance or party ideology, Dassonneville and Hooghe (2017) argue that economic context affects vote switching more broadly. When the economy is doing well, voters tend to stick with their previous party; during downturns, they are more likely to abandon it. Disillusioned voters may turn to emerging parties (Powell and Tucker, 2014). Weschle (2014) develops a more complex model where economic hardship leads some to switch, others to abstain, and still others to vote when they usually would not. Tuorto (2019) suggests that those returning to vote—especially in protest against traditional parties—are likely "protest abstentionists," whereas continued withdrawal from voting is more typical of the most socioeconomically disadvantaged. According to Weschle (2014), the dual effect of hardship can either increase or decrease aggregate turnout, depending on the balance between these opposing trends. This helps explain contradictory findings in the literature on the relationship between economic conditions and turnout (Blais, 2006).

However, the scarcity of suitable data remains a major obstacle. For instance, Dassonneville and Hooghe (2017) do not explore the hypothesized transition mechanisms directly. Instead, they analyze national-level "net" volatility—i.e., the difference in vote shares between two elections—and relate it to indicators like GDP growth, unemployment, and inflation. Yet net volatility underestimates "true" volatility at the individual level, potentially obscuring significant relationships (Dejaeghere and Dassonneville, 2017). This limitation hampers our understanding of which parties gain or lose from increased volatility and which factors drive gains and losses.

*The community level of analysis*

Economic shocks—such as the 2008 financial crisis—and structural transformations linked to technological change and market globalization have amplified the gradual dynamics mentioned earlier, deepening divisions between social groups. Globalization and technological advancement have disadvantaged less-skilled workers while benefiting more-skilled ones. Territorial disparities have also widened considerably: urban areas rich in human capital have profited from these transformations, while older industrial and rural areas have experienced decline, severely impacting local communities—and, consequently, the geography of voting.

This growing spatial inequality has generated an extensive literature on contextual effects, based on the assumption that voting decisions are shaped not only by individual preferences or social affiliations, but also by the local socio-economic environment. Johnston and Pattie (2006) argued early on that voters respond to local issues and to their perceptions of which parties best represent territorial interests. The uneven geographical consequences of economic change have prompted many scholars to focus on local communities as the most relevant unit of analysis—especially when studying the success of right-wing populist parties in advanced democracies. For instance, Rogers (2014) emphasizes the role of community-level economic conditions—independently of individuals' personal hardship—in driving "communotropic" economic voting. Broz et al. (2021) similarly argue that when individuals' well-being is tied to their local economy, the latter becomes a crucial analytical concept. Rodriguez-Pose (2018) identifies the roots of populist voting in the frustrations of people living in "left-behind" areas characterized by persistent poverty and economic stagnation. Bayerlein (2022) identifies spatial inequality induced by globalization as a key driver of right-wing populist support and suggests that public goods provision may mitigate its impact.

Cultural characteristics of local communities—such as resentment toward immigration or hostility to cosmopolitanism and multiculturalism—have also been identified as important explanatory factors for support of right-wing populist parties. The first is often intertwined with economic concerns (Hays et al., 2019; Bolet, 2020), while the second is more closely linked to geography, such as the urban-rural divide (Huijsmans, 2023). Arzheimer et al. (2024) show that the influence of contextual factors on far-right voting, driven by perceptions of local decline, is moderated by individuals' personal resources—particularly education—which can buffer the effects of unfavorable community conditions. Some authors, such as Bartle et al. (2017) and Song (2025), also suggest that spatial inequalities resulting from economic crisis and technological change are major contributors to increasing levels of abstention.

While most of these studies do not explicitly address vote switching, we believe that similar arguments apply. Johnston et al. (2005), in a study on household-level voting behavior, argue that "people who live together not only vote together, but also change their votes together." This reasoning may extend to broader community contexts. However, except our recent first attempt (see Bracalente et al. 2021), to date no studies have empirically examined how local factors affect vote switching—whether between parties or between voting and abstention. Instead, some studies have analyzed the geographical regularities in vote flows. Johnston and Hay (1983), using ecological estimates of voter transitions inside British constituencies, showed that Conservative and Labour loyalties between elections and vote switching between Labour and Conservative reflected an increasingly polarized geography. More recently, De Sio and Paparo (2014) used estimates of local-level transitions between electoral options to explore the persistence of traditional geopolitical patterns in Italy following the emergence of new populist actors like the Five Star Movement (M5S).[1]

**2.3 The proposed approach**

The literature reviewed above makes it clear that the main obstacle to empirical research on vote switching is the scarcity of adequate data. This helps explain why the determinants of vote change remain underexplored, despite electoral scholars recognizing that "switching parties between elections offers us an ideal window to observe general mechanisms of electoral decision making" (Dassonneville and Hooghe, 2017).

In line with several authoritative scholars, we argue that the lack of suitable data necessitates the use of ecological inference methods (Johnston and Hay, 1983; Johnston and Pattie, 2006). These methods allow us to estimate otherwise unavailable data—especially regarding the flow of votes from one party to another—and to appropriately analyze their determinants. Johnston and Pattie (2006) noted that, when used rigorously, ecological methods can yield results comparable to those obtained from survey-based studies. Nevertheless, in a comparative analysis of voter transitions between the 2006 and 2008 Italian general elections, Russo (2014) found that estimates based on the Goodman ecological method often diverged from corresponding survey-based estimates. However, given the serious recall biases in survey data, we argue that ecological estimates of vote switching may be less affected by systematic underestimation of electoral mobility than survey-based approaches. Furthermore, when transitions need to be estimated within small geographic areas, ecological inference becomes the only applicable method.

Based on these considerations, we propose a disaggregated approach to analyzing vote switching and its determinants, which includes the following steps:
1. *Definition of appropriate small areas* to serve as units of analysis for estimating voter transitions and measuring contextual predictors;
2. *Estimation of voter transition tables* for each small area using official polling station data and an advanced ecological inference method, followed by calculation of vote transition counts;
3. *Modeling the relationship between transition counts and contextual predictors* using multinomial logistic regression models.

The next section explores each of these stages in greater detail.

## 3. Small-area estimating of voting chances and their determinants

**3.1 Small area definition**

While it is widely recognized that local context matter, it remains unclear which specific geographic context is most suitable for studying voting behavior. Individual perceptions of what constitutes a relevant political context, combined with the spatial concentration of economic and social disadvantage, suggest the use of small areas. The appropriate size, however, depends on the purpose of the analysis.

Given the importance of local interaction effects, the context perceived as politically relevant at the individual level tends to be very small. Using mapping techniques, Wong et al. (2018) found that, in England and Wales, the average surface area of self-defined local contexts was about one-tenth the size of a parliamentary constituency. To analyze voting patterns in small towns versus rural areas in the 2020 U.S. presidential election, Mapes (2024) used precinct-level data (typically <1,000 voters), arguing that counties are too large to reveal meaningful behavior difference between small towns and rural areas. Other studies—such as those by McKey (2019) and Arzheimer et al. (2024)—used neighborhood-level data to examine perceptions of local decline. Bolet (2020) analyzed labour market competition between native and immigrant workers using municipal-level data, considering this the ideal scale to study radical right support. However, many relevant studies have used broader units. Abreu and Öner (2020), for example, employed parliamentary constituencies (~73,000 voters on average) to capture political mobilization effects in the Leave–Remain Brexit vote. Bayerlein (2022) examined the impact of spatial inequality on right-wing populism using county-level data in both the U.S. (~75,000 voters) and Germany (>150,000 voters).

---

[1] From a methodological perspective, it is also worth mentioning Puig and Giberna (2015), who developed an integrated method, combining cluster analysis and ecological inference, to estimate voter transitions across areas of Catalonia, aiming to minimize distortions caused by non-homogeneous voting behavior across polling stations.

For the methodological approach we propose, fairly small and homogeneous areas are required. This is because ecological inference can be biased if the assumption of voting behavior homogeneity across polling stations does not hold. Nevertheless, defining a set of zones requires balancing two other important conditions: (i) each zone must include enough polling stations to ensure reliable estimation of transition probabilities; (ii) the total number of zones must be large enough to allow estimation of multinomial regression models. To balance these competing needs, the small areas in our case study were based primarily on municipalities: individual mid-sized towns or groupings of smaller contiguous municipalities. For the largest city, sub-municipal zones were created by aggregating postal code areas.

### 3.2 Estimating voter transitions

We consider the estimation of voter transitions using ecological inference based on official polling station data to be, in general, a valid alternative to survey-based estimation. Moreover, within small geographic areas, voter transitions can only be estimated through ecological methods.

As just mentioned, ecological inference may be biased if the assumption of homogeneity in voting behavior across polling stations is violated. For an in-depth discussion of ecological fallacy, see Forcina and Pellegrino (2019). In brief, bias can arise, for example, when loyalty to a particular party is higher in polling stations where that party performed better in the previous election. However, such distortions are unlikely when the analysis is conducted within small, carefully defined areas.

The method used in this study to estimate voter transitions within each area is a modified version of the Brown and Payne (1986) model, as described in Forcina et al. (2012). Unlike the still commonly used Goodman model—which applies unweighted least squares—the approach adopted here relies on an asymptotic likelihood approximation and introduces a variance function adjustment. This method yields more efficient estimates and ensures that predicted transition probabilities fall between 0 and 1, without requiring post-hoc corrections. Once a table of estimated transition probabilities is available, one can compute the expected distribution of votes in the new election for each polling station and assess the fit between observed and predicted values. The algorithm also provides an estimate of the variance matrix on the logit scale, which can be used to derive standard errors for the estimated transition probabilities in each cell. For each small area, the conditional transition probabilities—i.e., those associated with voters' choices in the previous election—can be transformed into a matrix showing the expected number of votes shifting from each origin to each destination. These matrices, consistent with observed marginal totals (rows and columns), are produced for all areas and serve as the foundation for modeling the effects of contextual covariates on voter transitions. Additionally, volatility index at both the small-area and regional level can be calculated, the latter by aggregating the small area transition count tables.

### 3.2 Estimating the relationship between transitions and covariates

In principle, the logits of transition probabilities can be modeled as functions of covariates measured at the polling station level. However, very few contextual variables are available at that micro-geographic scale. Moreover, our findings suggest that such variables are often not the most relevant. For these reasons, we consider the use of the previously defined small-area units to be a more appropriate solution.

Regarding the factors affecting vote switching, it seems reasonable to assume that the local contextual and compositional characteristics identified in the literature as drivers of party support are also explanatory of vote transitions. For example, if the shift from traditional parties to populist ones is driven by dissatisfaction with the former, such shifts should be more frequent in areas most affected by structural change and economic crisis. Similarly, if the transition from voting to abstention signals general disillusionment with the political system, it too is likely to be more prevalent in those areas—though not necessarily triggered by the same specific factors. The covariates selected for our analysis reflect these assumptions.

As for the statistical approach, we estimate the relationships between vote transitions and local characteristics using multinomial logistic regression models fitted via maximum likelihood. The analysis is based on the collection of count tables—one for each small area—along with a set of compositional and contextual covariates measured at the same level.

Two modeling strategies can be adopted: (1) For a given *origin* category (from the first election), we can model how the distribution of votes across *destination* options varies with small-area covariates; (2) Conversely, for a given *destination* category (in the second election), we can model how the distribution of votes across *origins* varies with the same covariates. In both cases, the dependent variable is a vector of counts, and the estimated parameters include, for each non-reference option, an intercept and a set of coefficients capturing the effects of the covariates.

Further methodological details are provided in Appendix A, including: (i) a brief formal description of the multinomial logistic model; (ii) the empirical strategy used to select final models; (iii) the procedure for transforming regression coefficients into numerical derivatives of transition probabilities (marginal effects); (iv) a residuals analysis used to identify contextual effects not captured by the covariates.

### 4. The case-study

The purpose of this application is to examine how geographic, economic, and cultural disadvantages in local communities influenced the main voter transitions that occurred between the 2018 and 2022 national elections in the Italian region of Umbria.

The two primary destinations analyzed were:
- *Fratelli d'Italia (FdI)*, a nationalist right-wing party whose vote share increased from 5% to 31%, and
- *Abstention*, including blank and invalid ballots, which also rose significantly.

Umbria is particularly suitable for testing the proposed methodological approach. It was the first region in Italy's so-called "red zone"[2] to elect a right-wing populist party (Lega) to regional government in 2019. Despite its small size, Umbria includes a variety of local economic and social systems with different development trajectories. Though it lacks metropolitan areas, its urban fabric consists of medium-sized towns and small municipalities, along with substantial rural and peripheral zones. The region was also severely impacted by the financial crisis and deindustrialization, which negatively affected low-skilled employment and increased poverty.

**4.1 Construction of zones and aggregation of parties**

Umbria has approximately 700,000 voters distributed across about 1,000 polling stations. For this study, the region was divided into *nineteen small zones*: two were created by splitting the regional capital into urban and suburban sectors; five were defined as individual municipalities with at least 30,000 residents; the remaining twelve were formed by aggregating small, contiguous municipalities. Each zone contains an average of 37,000 voters and about 53 polling stations.

---

[2] The term "red zone," or "red regions," is often used to describe historically left-leaning areas such as Emilia-Romagna, Tuscany, Umbria, and the Marche.

To focus on the most politically relevant and numerically significant transitions, voting options with relatively low support were aggregated. These aggregations appear in the rows of Table 1, which presents, for each origin category, both the estimated transition counts at the regional level (upper part) and the corresponding row percentages (lower part). To analyze vote flows toward FdI and abstention, we used multinomial models for outgoing transitions. The analysis focuses on the loss of support from two populist parties, *Lega* and *Movimento 5 Stelle (M5S)*, two traditional pro-EU parties, *Partito Democratico (PD)* and *Forza Italia (FI)*, and the portion of *2018 abstainers* returned to vote in 2022.

Table 1 – Electoral transitions and their magnitude at the regional level (Umbria 2018-2022) [a,b]

| 2018 | 2022 | | | | | | Total |
|---|---|---|---|---|---|---|---|
| | M5S-OL | PD | OCL | FdI | Lega-FI-OCR | No vote | |
| | *Numbers of vote (thousands)* | | | | | | |
| M5S | 50.2 | 5.7 | 3.6 | 16.6 | 16.1 | 38.2 | 130.4 |
| PD | 5.6 | 69.6 | 9.9 | 6.8 | 3.9 | 23.9 | 119.7 |
| FI | 0.2 | 0.2 | 9.1 | 17.3 | 22.5 | 7.8 | 57.1 |
| Lega | 0.7 | 0 | 0.5 | 59.3 | 22 | 15.9 | 98.4 |
| No vote | 11 | 5.1 | 6 | 13.4 | 7.7 | 133.5 | 176.7 |
| | *Percentages on row totals* | | | | | | |
| M5S | 38.5 | 4.4 | 2.8 | 12.7 | 12.3 | 29.3 | 100.0 |
| PD | 4.7 | 58.1 | 8.3 | 5.7 | 3.3 | 20.0 | 100.0 |
| FI | 0.4 | 0.4 | 15.9 | 30.3 | 39.4 | 13.7 | 100.0 |
| Lega | 0.7 | 0.0 | 0.5 | 60.3 | 22.4 | 16.2 | 100.0 |
| No vote | 6.2 | 2.9 | 3.4 | 7.6 | 4.4 | 75.6 | 100.0 |

[a] *Party's acronyms: see Appendix B;*
[b] *Highlighted in gray: the outgoing transitions analyzed in an aggregated manner*

## 4.2 The covariates

The covariates and their component variables are listed in Table 2, along with summary statistics. These variables were selected to capture the main dimensions of *geographic*, *economic*, and *cultural disadvantage* affecting local communities, as identified in the literature as potential drivers of recent voting shifts. The underlying hypotheses behind the selection of covariates are as follows: vote switching in favor of the nationalist right-wing party and increased abstention are more likely to occur in areas that are:

- More *rural and peripheral*, relative to urban centers;
- *Slower to recover* from the effects of the post-2008 financial crisis and more exposed to unemployment;
- Characterized by a *higher share of low-skilled workers*, who were more severely affected by structural and technological changes;
- Marked by *lower education levels, weaker social capital, and greater distrust toward the EU*;
- Exhibiting a *weaker left-wing political tradition*, which might influence rightward shifts.

Some covariates were constructed as the average of multiple standardized elementary indicators (these are highlighted in grey in Table 2). All covariates were standardized to have zero mean and unit variance, except for the political tradition variable, which is dichotomous. Where necessary, the sign of the standardized variable was inverted (i.e., multiplied by –1) so that higher values consistently indicate greater disadvantage.

Table 2 – Covariates and descriptive statistics of the elementary variables [a]

| Elementary variables by categories | Covariates | Descriptive statistics | | | |
|---|---|---|---|---|---|
| | | Mean | St. Dev. | Max | Min |
| *Geographical* | | | | | |
| X1 - Percentage of population in municipalities and outskirts with up to 5,000 inhabitants | geog | 60.0 | 23.9 | 93.0 | 0.0 |
| X2 - Distance from the provincial capital (in minutes by car) | | 33.1 | 16.2 | 58.9 | 0 |
| *Economic* | | | | | |
| X3 - Percentage of annual income earners up to 15,000 euros, 2021 | income | 42.6 | 2.6 | 47.3 | 38.2 |
| X4 - Per capita income from employment, 2021 | skill | 19.0 | 0.9 | 21.5 | 17.4 |
| X5 – Rate of unemployment, 2021 | unempl | 5.6 | 0.9 | 7.5 | 4.3 |
| X6 - Index number of income variation from personal taxes, 2008-2021 | recovery | -1.0 | 4.2 | 8.4 | -5.4 |
| X7 - Variation 2008-2021 in the percentage of income earners up to 15,000 euros | | -3.2 | 2.2 | 1.4 | -6.6 |
| *Cultural* | | | | | |
| X8 - Percentage of qualifications up to lower secondary school, 2021 | educ | 45.3 | 4.0 | 50.3 | 32.9 |
| X9 - Institutions with volunteers per 100 inhabitants, census 2011 | | 0.6 | 0.1 | 0.9 | 0.5 |
| X10 - Volunteers per 100 inhabitants, census 2011 | ksoc | 11.9 | 2.3 | 16.3 | 7.8 |
| X11 - Volunteer organizations registered in the regional register per 1,000 inhabitants, 2021 | | 0.8 | 0.2 | 1.2 | 0.3 |
| X12 - Average percentage of voters in the constitutional referendums, 2016 and 2020 | | 61.2 | 2.8 | 65.8 | 55.4 |
| X13 - Increase in the % of abstainers from 2018 national elections to 2019 EP elections | eutrust | 9.6 | 5.4 | 19.8 | 0.2 |
| X14 - Political tradition index: 1 left-wing; 0 otherwise | lefttrad | | | 1 | 0 |

[a] *highlighted in gray: elementary variables aggregated into a single covariate*

A few clarifications on variable construction:
- *Geographic disadvantage* was measured using indicators of *rurality* and *peripherality*. Although representative of partially distinct geographical aspects, due to their high correlation ($r = 0.55$) these were combined into a single geographic index (*geog*) in order to reduce number of covariates and multicollinearity.
- The *prevalence of low-skilled employment* was estimated using the inverse of per capita wage income, based on the assumption that lower wages correlate with lower worker qualifications.
- Because *income*, *education*, and *skill level* were all highly correlated with the geographic variable *geog* (with *r* between 0.79 and 0.81), and of each other as a consequence, they were *residualized* by regressing each on *geog*. The resulting residuals—*skill, income,* and *educ*—are uncorrelated with *geog* and only weakly correlated among themselves (*r* values between 0.28 and 0.40, compared to 0.74–0.78 for the original variables) They represent the specific contribution of each factor, independent of common geographic factor.
- *Slow economic recovery* was measured using a combined indicator capturing each area's deviation from the national average in two respects: (i) percentage of recovery of pre-crisis income levels, and (ii) the change in the share of low-income earners.
- *Social capital (ksoc)* was constructed from four indicators: the presence of voluntary associations (from institutional and regional registries), volunteering participation rates, and turnout in the 2016 and 2020 constitutional referenda.
- *Distrust toward the EU (eutrust)* was indirectly measured as the increase in percentage of abstention in the 2019 European elections relative to the 2018 national elections.
- *Left-wing political tradition (lefttrad)* was coded as a dichotomous variable: zones where the Partito Comunista Italiano (*PCI*) received more than 1.5 times the votes of the Democrazia Cristiana (*DC*) in 1987 general elections (the last in which the two parties both participated) were coded as "1" (left-wing tradition), and "0" otherwise.

Apart from *recovery*, which remains correlated with *geog* ($r = -0.75$) as the two largest cities have been hit hardest by the crisis, all other covariates were sufficiently distinct (correlations between 0 and 0.45) to allow for independent effect estimation.

**4.3 The main results**

*Electoral volatility*

Before analyzing the factors influencing voter transitions, it is important to highlight the magnitude and composition of the observed electoral volatility, as well as its territorial patterns. Overall, the estimated level of electoral volatility was very high.[3] A majority of voters from the 2018 elections changed their choice by 2022: 39% switched to a different party; 19% abstained. For all major parties except the *PD* (58% of loyalty) total volatility exceeded 60%.

Figure 1 displays the territorial distribution of two components of volatility: volatility between parties was lowest (27%) in the western zones of the region—areas with the strongest left-wing political traditions—and peaked (near 50%) in medium-small cities in the south, which were heavily impacted by industrial decline; volatility toward abstention showed a mirror-image pattern: it was highest where party switching was lowest and vice-versa. The negative correlation between these two volatility measures ($r = -0.42$) suggests that for local communities switching to other parties and dropping out of the electoral process are substitute behaviors, not complementary. In other words, when dissatisfaction arises, communities that more increase switching parties tend to less increase switching towards abstention and vice-versa—not to more increase both.[4]

Figure 1 – Umbria maps of between parties and towards abstention zonal volatility measures

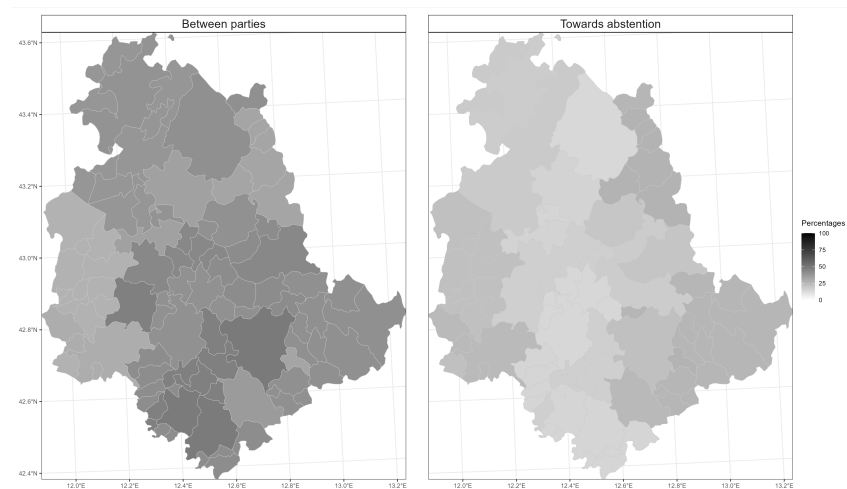

*Estimate models*

The final multinomial models for outgoing transitions are presented in Appendix C (Table C.1). Model fit varies by origin party, with explained deviance ranging from 48.7% for M5S to 66.5% for PD.[5] In each model, loyalty to the original choice is treated as the reference category in the logit

---

[3] The zonal and regional volatility indexes were calculated by also including two minor origin categories (small left-wing and small right-wing parties) non reported in Table 1. Both indexes were calculated as the percentage of total vote switching relative to all votes cast for parties in the first election.

[4] In the most marginal mountainous area (southeastern part of the region), the two forms of electoral volatility appear to combine.

[5] The residual analysis of the initial models for outgoing transitions from FI and M5S revealed two positive outliers in transitions toward FdI and PD, respectively (see Figure C1 in Appendix C). These outliers were due to the presence of

specification. While this choice affects the interpretation of coefficients, it does not alter model fit or the significance of explanatory variables. For interpretability, Tables 3 and 4 present marginal effects ($\Delta p$) only for statistically significant covariates.

*Transitions toward the right-wing nationalist party (FdI)*

Several covariates reflecting geographic, economic, and cultural disadvantage, as well as weaker left-wing political tradition, were found to promote switching toward FdI—though effects varied by origin. Among previous abstainers, the strongest predictors of switching to FdI (all significant at the 1% level) were rurality/peripherality, slow post-crisis recovery, lower skill levels, and lower education ($\Delta p$ ranging from 0.049 to 0.12). In contrast, poverty of income and low social capital were associated with continued abstention, not re-engagement via FdI. This supports the idea that the right-wing opposition party attract "reactivated" voters from communities hurt by the post-2008 crisis and economic restructuring—in small towns with low skill levels and education. Meanwhile, communities poorer of social capital and with prevalence of low-income earners tend to discourage participation.

Table 3 – Marginal effects of significant covariates on transition probabilities toward FdI and No vote[a, b]

| Origin | Covariates | | | | | | | | |
|---|---|---|---|---|---|---|---|---|---|
| | Geographic | Economic | | | | Cultural | | | |
| | geog | recovery | skill | income | unempl | educ | ksoc | eutrust | lefttrad |
| *Outgoing transitions towards FdI* | | | | | | | | | |
| M5S | 0 | 0 | 0 | 0 | 0 | 0 | *0.021* | **0.044** | -0.033 |
| PD | 0 | -0.013 | 0.019 | *-0.008* | 0 | **-0.011** | **0.015** | 0 | **-0.013** |
| FI | 0 | *0.082* | 0 | 0 | 0 | *0.083* | 0 | 0 | 0 |
| Lega | 0 | 0.097 | 0 | **0.122** | 0 | **-0.123** | 0.078 | 0 | 0 |
| No vote | **0.115** | **0.12** | **0.049** | **-0.069** | 0 | **0.071** | -0.101 | 0 | 0 |
| *Outgoing transitions towards No vote* | | | | | | | | | |
| M5S | 0 | 0 | 0 | 0 | 0 | 0 | 0 | **0.09** | 0 |
| PD | 0 | 0 | 0 | 0 | 0 | 0 | 0 | 0 | *-0.027* |
| FI | 0 | 0 | 0 | *-0.100* | 0 | 0.154 | *-0.130* | 0 | 0 |
| Lega | 0 | **0.051** | 0 | 0 | 0 | **-0.069** | 0.051 | 0 | *-0.031* |

[a] Probability variation: see Appendix A; Party's acronyms: see Appendix B; Covariate legends: see Sections 4.2
[b] Bold: p-vaue < 0.01; Italic: p-value < 0.08

Outflows from pro-EU parties was also partly interpretable as a locally based economic voting. Those from FI were associated with slow economic recovery and low education, while those from PD with low skill jobs. However, the latter were influenced more by cultural factors (political tradition and low social capital) than by economic ones. Interestingly, lower levels of education and income had a negative effect, implying that the PD's popular and disadvantaged strongholds resisted the extreme vote switching towards FdI, as also testified by the negative contribution of the left-wing political tradition. Similarly, one can probably interpret the surprising negative effect of the greater intensity of the post-2008 economic crisis (and therefore the delay in recovery). In fact, the crisis mainly affected the two largest cities of the region, where the PD is also more rooted.

---

local candidates from those parties in zones 2 and 17 during the 2022 national elections. To account for these specific local effects, the models were re-estimated, including statistically significant dummy variables for the two zones.

Conversely, among Lega voters switching to FdI was driven by economic factors, as slow economic recovery and lower-income earners, but mitigated by low education, which acted as a protective factor against political realignment. Among M5S voters, transitions to FdI were also mainly shaped by cultural factors, as weaker left-wing tradition, anti-EU attitudes, and low social capital. Across origins, unemployment and low income showed either null or negative effects on switching to FdI—supporting Inglehart and Norris's (2016) claim that such variables are poor predictors of right-wing populist support.[6]

*Transitions towards abstention*

The effects of local disadvantages on transitions toward abstention differ markedly from those found for transitions toward *Fratelli d'Italia (FdI)*. Notably, social and cultural factors appear to have played a much stronger role in driving abstention, while economic factors had a more limited or inconsistent effect. Again, Lega voters are the exception, being pushed by the slower recovery from the economic crisis to also abstain, while low education acted as protective factor also against political disengagement; other social factors, as low social capital, instead increased the shift to abstention. Even with regard to the propensity to abstain, the negative attitude towards the EU and the poor education are confirmed as factors of vulnerability for M5S and FI, respectively, while for PD the left-wing tradition remain a protection factor.[7] Overall, these findings suggest that choice of switching towards abstention—unlike shift to the right-wing opposition party—is not mainly fostered by economic hardship, but rather by a broader erosion of civic engagement depending on cultural disadvantages and lack of social cohesion.

However, the almost null effects of rurality–peripherality factor remains to be explained. This unexpected result could be symptomatic of changing tendencies regarding abstention, in which two antithetic components seem now coexist: the apathy towards politics, more present in rural and marginal areas, and the new tendence to abstain from voting of young generations (mainly urban), also because it is considered less effective than other forms of political engagement (Sloam, 2016, among others).

## 5. Discussion and conclusions

This paper has presented a methodological approach for estimating vote transitions and analyzing their determinants in the absence of reliable survey data. The proposed method is based on the use of official polling station data to estimate the number of transitions within small homogeneous areas and to explain those transitions using multinomial logistic regression models, with covariates measured at the same geographical level. The application focused on vote shifts in the 2018–2022 general elections in the Italian region of Umbria, with particular attention to transitions toward the nationalist right-wing party that won the elections and toward the rising abstention.

The results confirm that both types of transitions were influenced by contextual disadvantages in the local communities, but with distinct patterns of association. Transitions toward the right-wing party were associate with economic disadvantage, particularly in the case of previous abstainers and Lega voters. They also appeared to be associated with cultural disadvantage—especially among M5S and PD voters—with variables such as weak left-wing political tradition, low social capital, and anti-EU sentiments showing positive and significant effects for one or the other of the two parties. These findings align with previous literature emphasizing the relevance of territorial inequalities and cultural resentment in explaining right-wing populist support, together with the persistent opposite effect of left-wing political tradition. However, they also suggest that the specific channels through which these factors operate may differ depending on the political origin of the mobile voters.

---

[6] The positive impact of prevalence of lower-income earners on the transitions out of the (far-right) Lega is also in line with this statement. For further discussion of the effects of contextual unemployment, see Sipma & Lubbers (2020).
[7] The negative effects of lower incomes and lack of social capital on the transition out of FI, besides being less statistically significant, are also controversial in their interpretation.

Transitions toward abstention showed a different profile, in some ways opposite to that observed for transitions toward FdI. They were mainly driven by social and cultural disadvantages, while the indicators of economic hardship had null or even negative effects. For some origins, lack of social capital, low education, and anti-EU sentiments emerged as significant positive predictors. This suggests that, rather than being two expressions of the same form of dissatisfaction, abstention and voting for a right-wing opposition party respond to different contextual triggers.

Overall, the findings support the idea that geographic, economic and cultural context matters, and they demonstrate that a disaggregated ecological approach can be effectively used to analyze the mechanisms of vote switching. The results also show that electoral abstention and support for populist parties should not be lumped together, as they are associated with different drivers.

The main limitation of the case study is the small size of the analyzed region. The small total amount of polling stations available didn't allow to construct a sufficiently large number of homogeneous areas with a sufficiently large number of polling stations inside. This constrained the size of transition tables, and consequently the disaggregation of transitions flows, and made it harder to estimate their relationships with local factors. Furthermore, the small number of polling stations inside some of the zones, while reducing the bias inherent in ecological inference, increased the variance and hence the standard error of the voter transition estimates. Expanding the study area could strengthen the findings, giving more solidity to the exemplificative application of the proposed approach.

From a methodological perspective, this paper highlights the potential of combining ecological inference with contextual modeling to overcome the lack of reliable individual-level data. By estimating vote transitions within small areas and analyzing them with appropriate regression models, it is possible to gain insight into the mechanisms of electoral change that would otherwise remain hidden. This is particularly relevant in contexts where survey data are affected by severe nonresponse and recall errors, or where sample sizes are too small to allow local-level analysis.

Future applications of this approach could help explore the spatially differentiated dynamics of recent electoral changes in other regions or countries. They could also examine how different types of local disadvantages interact in shaping the electoral behavior—both in terms of party switching and voter (re)mobilization or demobilization. In this way, the method could contribute to a better understanding of the local foundations of electoral change.

**References**

Abreu, M., Öner, Ö. (2020). Disentangling the Brexit vote: The role of economic, social and cultural contexts in explaining the UK's EU referendum vote. *Environment and Planning A: Economic and Space*, 52(7), 1434-1456. https://doi.org/10.1177/0308518X20910752

Adida, C., Gottlieb, J., Kramon, E., McClendon, C., (2019). Response Bias in Survey Measures of Voter Behavior: Implications for Measurement and Inference. *Journal of Experimental Political Science*, 6 (3), 192–198. https://doi.org/10.1017/XPS.2019.9

Arzheimer et al. (2024). How local context affects populist radical right support: A cross-national investigation into mediated and moderated relationships. *British Journal of Political Science*, 54(4), 1133-1158. https://doi.org/10.1017/S0007123424000085

Bayerlein, M. (2022). Regional deprivation and populism: Evidence from Germany and the U.S. *Kiel Working Paper*, No. 2231, Kiel Institute for the World Economy, Kiel. http://www.ifw-kiel.de/

Bartle, J., Birch, S., Skirmuntt, M. (2017). The local roots of the participation gap: Inequality and voter turnout, *Electoral Studies*, 48, 30-44. https://doi.org/10.1016/j.electstud.2017.05.004

Bernstein, R., Chadha, A., Montjoy, R. (2001). Overreporting voting. Why it happens and why it matters. *Public Opinion Quarterly*, 65, 22–44. doi: 10.1086/320036

Bodet, M.A., Laflamme, L., Brie, E., Ouellet, C. (2025). Exit Polls in Canada: A Methodological Note. *Canadian Journal of Political Science*. https://doi.org/10.1017/S0008423924000398

Bolet, D. (2020). Local labour market competition and radical right voting: Evidence from France. *European Journal of Political Research,* 59, 817–841, doi: 10.1111/1475-6765.12378.

Bracalente, B., Forcina, A., Falocci, N. (2021). Fattori di contesto e flussi elettorali: un'analisi empirica del caso Umbria, *Polis*, 35(3), 379-412. DOI: 10.1424/102288

Brown, P.J., Payne, C.D. (1986). Aggregate data, ecological regression, and voting transitions. *Journal of the American Statistical Associa*tion, 81, 452–460. https://doi.org/10.2307/2289235

Broz, J.L., Frieden, J., Weymouth, S. (2021). Populism in Place: The Economic Geography of the Globalization Backlash. *International Organization*. 1-31. DOI: 10.1017/S0020818320000314

Dalton, R.J. (1984), Cognitive Mobilization and Partisan Dealignment in Advanced Industrial Democracies. *The Journal of Politics*, 46 (1), 264-284. https://doi.org/10.2307/2130444

Dassonneville, R. (2012). Electoral Volatility, Political Sophistication, Trust and Efficacy: A Study on Changes in Voter Preferences during the Belgian Regional Elections of 2009. *Acta Politica*, 47(1), 18–41. https://doi.org/10.1057/ap.2011.19

Dassonneville, R., Blais, A., Dejaeghere, Y. (2015). Staying With the Party, Switching or Exiting? A Comparative Analysis of Determinants of Party Switching and Abstaining. *Journal of Elections, Public Opinion and Parties*, 25 (3), 387-405. http://dx.doi.org/10.1080/17457289.2015.1016528

Dassonneville, R., Hooghe, M. (2017). Economic Indicators and Electoral Volatility Economic Effects on Electoral Volatility in Western Europe, 1950-2013. *Comparative European Politics,* 15(6), 919-943 DOI:10.1057/cep.2015.3

Dejaeghere and Dassonneville (2017). A comparative investigation into the effects of party-system variables on party switching using individual-level data. *Party Politics*, 23(2), 110-123. DOI: 10.1177/1354068815576294

Dijkstra, L., Poeleman, H., Rodrìguez-Pose, A. (2020) The geography of EU discontent. *Regional Studies*, 54(6), 737-753. https://doi.org/10.1080/00343404.2019.1654603

Durand, C., Deslauriers, M., Valois, I. (2015). Should Recall of Previous Votes Be Used to Adjust Estimates of Voting Intention? Survey Insights: Methods from the Field. Weighting: Practical Issues and 'How to' Approach. https://surveyinsights.org/?p=3543

Fieldhouse, E., Green, J., Evans, G., Mellon, J., Prosser, C., Schmitt, H., van der Eijk, C. (2020). Electoral Shocks. Oxford University Press. https://doi.org/10.1093/oso/9780198800583.001.0001

Forcina, A., Gnaldi, M., Bracalente, B. (2012). A Revised Brown and Payne Model of Voting Behaviour Applied to the 2009 Elections in Italy. *Statistical Methods & Applications*, 21 (1), 109-119. https://doi.org/10.1007/s10260-011-0184-x

Forcina, A., Pellegrino, D. (2019). Estimation of Voter Transitions and the Ecological Fallacy. *Quality & Quantity*, 53 (4), 1859-1874. https://doi.org/10.1007/s11135-019-00845-1

Guriev, S., Papaioannou, E., 2022. The Political Economy of Populism. Journal of Economic Literature. 60 (3), 753–832. DOI: 10.1257/jel.20201595

Huijsmans T. (2023). Why some places don't seem to matter: Socioeconomic, cultural and political determinants of place resentment. *Electoral Studies*, 83, 102622. https://doi.org/10.1016/j.electstud.2023.102622

Inglehart, R. (2008). Changing values among western public from 1970 to 2006. *West European Politics*, 31(1-2), 130-146. https://doi.org/10.1080/01402380701834747

Inglehart, R., Norris, P. (2016). Trump, Brexit, and the Rise of Populism: Economic Have-Nots and Cultural Backlash. *Harvard Kennedy School Working Paper*. RWP16-026. https://www.hks.harvard.edu

Johnston R.J., Hay, A.M. (1983). Voter Transition Probability Estimates: An Entropy-Maximizing Approach. *European Journal of Political Research*, 11(1), 93-98. https://doi.org/10.1111/j.1475-6765.1983.tb00045.x

Johnston, R.J., Pattie, C. (2006). Putting Voters in Their Place: Geography and Elections in Great Britain. Oxford University Press. https://doi.org/10.1093/acprof:oso/9780199268047.001.0001

Johnston, R. J., Jones, K., Propper, C., Sarker, R., Burgess, S., Bolster, A. (2005) A missing level in the analysis of British voting behaviour: the household as context as shown by analyses of a 1992–1997 longitudinal survey. *Electoral Studies*, 24, 201–25. https://doi.org/10.1016/j.electstud.2004.04.002

Mapes, J. (2024). Using big data to study small places: Small town voting patterns in the 2020 U.S. presidential election, *Growth & Change*, e12730. https://doi.org/10.1111/grow.12730

McAllister, I, Quillan S. (2022). Vote overreporting in national election surveys: a 55-nation exploratory study. *Acta Politica*, 57, 529–547. https://doi.org/10.1057/s41269-021-00207-6

McKay, L. (2019). 'Left behind' people, or places? The role of local economies in perceived community representation. *Electoral Studies*, 60, 1–11. https://doi.org/10.1016/j.electstud.2019.04.010

Pavia, J.M., Badal, E., García-Cárceles, B. (2016). Spanish Exit Polls: Sampling error or nonresponse bias? *Revista Internacional de Sociologica*, 74 (3), 1–15. http://dx.doi.org/10.3989/ris.2016.74.3.043

Powell, E.N., Tuker, J.A. (2014) Revisiting electoral volatility in post-communist countries: New data, new results and new approach. *British Journal of Political Science*, 44(1), 123-147. https://doi.org/10.1017/S0007123412000531

Rodríguez-Pose, A. (2018). The revenge of the places that don't matter (and what to do about it). *Cambridge Journal of Regions, Economy and Society*, 11(1), 189-209. https://doi.org/10.1093/cjres/rsx024

Rogers J. (2014). A communotropic theory of economic voting. *Electoral Studies*, 36, 107-116. https://doi.org/10.1016/j.electstud.2014.08.004

Sloam, J. (2016). Diversity and voice: The political participation of young people in the European Union. *British Journal of Politics & International Relations*, 18(3), 1-26. https://doi.org/10.1177/1369148116647176

Song, B.K., Woo Chang Kang (2025). Inequality, local wealth, and electoral politics. *European Journal of Political Economy*, 86, 102617. https://doi.org/10.1016/j.ejpoleco.2024.102617

Tuorto, D. (2019). I non rappresentati. La galassia dell'astensione prima e dopo il voto del 2018. *Teoria Politica*, 8, 263-273. http://journals.openedition.org/tp/361

Waldahl, R., Aardal, B. (2000). The Accuracy of Recalled Previous Voting: Evidence from Norwegian Election Study Panels, *Scandinavian Political Studies*, 23(4), 373-389. https://doi.org/10.1111/1467-9477.00042

Weschle, S. (2014). Two types of economic voting: How economic conditions jointly affect vote choice and turnout. *Electoral Studies*, 34, 39-53. https://doi.org/10.1016/j.electstud.2013.10.007

Wong, C., Bowers, J. Rubenson, D., Fredrickson, M., Rundlett, A. (2020). Maps in People's Heads: Assessing A New Measure of Context. *Political Science Research and Methods*, 8(1), 160-168. https://doi.org/10.1017/psrm.2018.51

## Appendix A - Methodological details

Let $n_{zj}$ represent the number of votes (estimated through ecological inference) in zone $z$ that moved from the party of interest to party $j$. Let $k$ represent the number of voting options considered; the model for outgoing transitions assumes that the vector with elements $n_{z1}, ..., n_{zk}$, where $z$ indicates the zone and $k$ the reference option, follows a multinomial distribution with a total of $n_z$, the number of votes of the party considered in zone $z$ in the previous election, and a probability vector $p_{z1}, ..., p_{zk}$. Assuming that the logits are a linear function of the covariates, *the multinomial logistic model* for outgoing transitions assumes that

$$\log \frac{p_{zj}}{p_{zk}} = \beta_{0j} + \sum_v x_{zjv} \beta_{jv}$$

where $v$ indexes the different covariates and $j$ the exit option. The parameters to be estimated by maximum likelihood include a different intercept for each exit option $j$ except reference option $k$ and, for each exit category and each covariate, a parameter $\beta_{jv}$ measuring the effect of covariate $v$ on exit option $j$. To simplify interpretation, the covariates are expressed as deviations from their mean, so that $\beta_{0j}$ is the value of the logit when all covariates are equal to zero. A similar scheme can be defined for the logistic model applied to incoming transitions.

To facilitate assessing *the effect of covariates on transition probabilities*, in addition to the regression coefficients, which are expressed on the logistic scale, numerical derivatives were also computed. Let $p_j(x)$ be the probability of transition to (or from) option $j$ based on the selected logistic model estimated in a zone where the covariates equal $x$; let $t$ be an arbitrarily small value and $x_{v,t}$ the vector of covariates where all elements are set to 0 except for covariate $v$, which is assigned the value $t$ (set at 0.00001 in the case study); we define the probability variation (*marginal effects*) for covariate $v$ and option $j$ as

$$\nabla p_j = \frac{p_j(x_{v,t}) - p_j(\mathbf{0})}{t}$$

However, it should be noted that these variations isolate the effect of one covariate on a single exit (or entry) without accounting for indirect effects when the covariate in question affects other exits/entries. Moreover, since the derivative of the logit function tends toward 0 when $p_j$ is close to 0 or 1, the effects on probability are also dampened near the extremes.

Regarding *the choice of the logistic regression models*, the difficulty in formulating theoretical hypotheses on the determinants of the many transitions analyzed suggested the adoption of an empirical procedure for selecting the models. As a starting point, we constructed a broad set of covariates relating to the main geographical, economic, and cultural aspects identified in the literature as possible determinants of the most significant voting changes in recent years, whose effects for some of them may, however, prove irrelevant in the specific case study. For each exit or entry of interest, the optimal model was selected, by starting from the model where all covariates (nine in the case study) affect each $p_{zj}$, removing, in a step-wise manner, those where the value of the ratio $z_{jv} = \hat{\beta}_{jv}/[Var(\hat{\beta}_{jv})]^{1/2}$ was less than a certain threshold (assumed to indicate a clear lack of statistical significance); increasing this threshold every time a reduced model was estimated by removing the covariates whose $z_{jv}$ did not exceed the threshold.

To identify zones where the explanatory capacity of the covariates in relation to voting behavior is inadequate—indicating the presence of *context effects not captured by the covariates*— we used the following procedure: i) the difference between the transition probabilities estimated via ecological inference ($n_{zij}/n_{zik}$) and those predicted by the regression model, that is $\hat{n}_{zij}/\hat{n}_{zik}$, were calculated; ii) the standard errors of these residuals were first approximated from the covariance matrix of the multinomial logistic regression model, assuming that the frequencies $n_{zij}$ follow a multinomial distribution, then corrected to take into account that the dependent variables, $n_{zij}$, are not observed but estimated through ecological inference; iii) the residuals were standardized dividing by their standard errors, and those outside the confidence intervals were considered indicative of the presence of context effects not captured by the explanatory variables.

**Appendix B – Details on the Italian parties, acronyms and party aggregations**

*Fratelli d'Italia* (FdI) – Right-wing nationalist party, won the 2022 elections, leads the Italian government; *Partito Democratico* (PD) – Centre-left pro-EU party, lead or supported several Italian governments until 2022; *Movimento 5 Stelle* (M5S) – Born as an anti-system populist movement, after 2018 it leaded the Italian government in alliance first with the Lega and then with the PD;
*Lega* – Born as an ethno-regionalist party (Lega Nord), it transformed into a populist right-wing, anti-Europe party; *Forza Italia* (FI) – Centre-right pro-EU party, long in government also led by its founder Berlusconi;
*Others Left* (OL) – Includes *Alleanza Verdi-Sinistra* and others radical left-wing parties;
*Others Centre-Left* (OCL) – Includes small moderate Centre-left parties: Italia Viva, Azione and +Europa;
*Others Centre-Right* (OCR) – Includes smaller, Centre-right independent parties.

# Appendix C - Tables and Graphics

Table C1 – Multinomial logit models for outgoing transitions toward FdI and No vote [a]

|  | Destination | Coeff. | Geographic | Economics | | | | Cultural | | | |
|---|---|---|---|---|---|---|---|---|---|---|---|
|  |  |  | geog | recov | skill | income | unempl | educ | ksoc | eutrust | lefttrad |
| Transitions out of M5S (% dev. 48.7) | FDI | b | 0 | 0 | -0.25 | 0 | 0 | 0 | 0.24 | 0.48 | -0.38 |
|  |  | z |  |  | -1.54 |  |  |  | 1.96 | 3.06 | -2.87 |
|  | No vote | b | 0.28 | 0.18 | 0 | 0 | 0 | 0 | 0 | 0.39 | 0 |
|  |  | z | 1.7 | 1.32 |  |  |  |  |  | 4.08 |  |
|  | PD-OCL | b | 0 | 0 | 0 | -0.7 | -0.42 | 0 | 0.44 | 0 | -0.35 |
|  |  | z |  |  |  | -3.5 | -2.51 |  | 2.69 |  | -2.10 |
|  | FI-Lega-OCR | b | -0.19 | 0 | 0 | 0.47 | 0 | -0.33 | 0.35 | 0.14 | 0 |
|  |  | z | -1.48 |  |  | 3.31 |  | -1.94 | 2.02 | 1.08 |  |
| Transitions out of PD (% dev. 66.5) | FDI | b | -0.69 | -0.86 | 1.26 | -0.53 | 0 | -0.76 | 0.98 | 0.43 | -0.89 |
|  |  | z | -1.04 | -2 | 3.15 | -1.75 |  | -2.64 | 3.68 | 1.45 | -3.36 |
|  | No vote | b | 0.11 | 0 | 0 | -0.17 | 0 | 0 | 0 | 0 | -0.17 |
|  |  | z | 1.35 |  |  | -1.64 |  |  |  |  | -1.76 |
|  | M5S-OL | b | 3.82 | 3.41 | -0.66 | -1.11 | -1.2 | 1.89 | 0 | 2.44 | -0.88 |
|  |  | z | 3.02 | 3.33 | -1.65 | -1.52 | -2.77 | 2.35 |  | 3.25 | -2.10 |
|  | OCL | b | 0 | 0 | 0.22 | 0.33 | -0.39 | -0.33 | 0 | -0.23 | 0 |
|  |  | z |  |  | 1.34 | 2.23 | -3.04 | -2.45 |  | -1.39 |  |
|  | FI-Lega-OCR | b | 0 | 6.41 | 6.25 | -6.94 | -7.34 | 5.69 | -7.27 | -7.63 | 0 |
|  |  | z |  | 2.78 | 2.89 | -2.91 | -2.73 | 2.61 | -2.68 | -3.06 |  |
| Transitions out of FI (% dev. 57.6) | FDI | b | 0 | 0.21 | 0 | 0 | 0 | 0.32 | 0 | -0.21 | 0.30 |
|  |  | z |  | 1.36 |  |  |  | 1.8 |  | -1.14 | 1.64 |
|  | No vote | b | 1.41 | 1.41 | 0.71 | -1.06 | -0.79 | 1.91 | -1.35 | 0 | 0 |
|  |  | z | 1.46 | 1.19 | 1.71 | -1.92 | -1.44 | 2.53 | -1.82 |  |  |
|  | M5S-OL-PD-OCL | b | 0 | 0 | 0.26 | -0.37 | 0.33 | 0.49 | -0.93 | 0 | 0 |
|  |  | z |  |  | 1 | -1.41 | 1.61 | 1.56 | -2.76 |  |  |
| Transitions out of Lega (% dev. 51.7) | FDI | b | 0 | 0.44 | -0.25 | 0.55 | -0.26 | -0.56 | 0.36 | 0 | 0 |
|  |  | z |  | 2.44 | -1.82 | 3.69 | -1.46 | -2.97 | 1.83 |  |  |
|  | No vote | b | 0 | 0.49 |  | 0.34 | 0 | -0.67 | 0.49 | 0 | -0.30 |
|  |  | z |  | 3.04 |  | 1.63 |  | -2.6 | 2.03 |  | -1.82 |
|  | M5S-OL-PD-OCL | b | 0 | 0 | 0 | 2.62 | -2.11 | 0.99 | 1.44 | 0 | 0 |
|  |  | z |  |  |  | 2.26 | -1.5 | 1.57 | 1.48 |  |  |
| Transitions out of No vote (% dev.62.9) | FDI | b | 2.04 | 2.12 | 0.87 | -1.21 | -0.58 | 1.26 | -1.79 | -0.31 | 0 |
|  |  | z | 3.61 | 2.58 | 2.8 | -3.09 | -1.45 | 2.68 | -3.25 | -1.47 |  |
|  | M5S-OL | b | 0 | 0 | 0.83 | -0.8 | 0.74 | 0.49 | -1.12 | 0 | -0.27 |
|  |  | z |  |  | 3.1 | -2.2 | 3.34 | 1.33 | -2.94 |  | -1.03 |
|  | PD | b | 1.22 | 3.27 | -1.23 | 1.05 | -1.21 | 0 | 1.57 | 0.67 | -1.17 |
|  |  | z | 1.38 | 2.1 | -1.89 | 1.59 | -1.79 |  | 2.09 | 1.71 | -2.64 |
|  | OCL | b | 1.39 | 2.22 | -0.99 | 0.5 | -0.51 | 0 | 0.73 | 0 | 0 |
|  |  | z | 2.81 | 2.68 | -2.21 | 1.27 | -1.37 |  | 1.84 |  |  |
|  | FI-Lega-OCR | b | 2.15 | 2.78 | 0 | -0.6 | -1.39 | 0.9 | -0.91 | 0.83 | -0.27 |
|  |  | z | 3.16 | 3.07 |  | -1.84 | -2.89 | 1.74 | -1.84 | 2.74 | -1.15 |

[a] *Party's acronyms: see Appendix B*

Figure C1 – Standardized discrepancies from the base transition models out of FI and M5S[a]

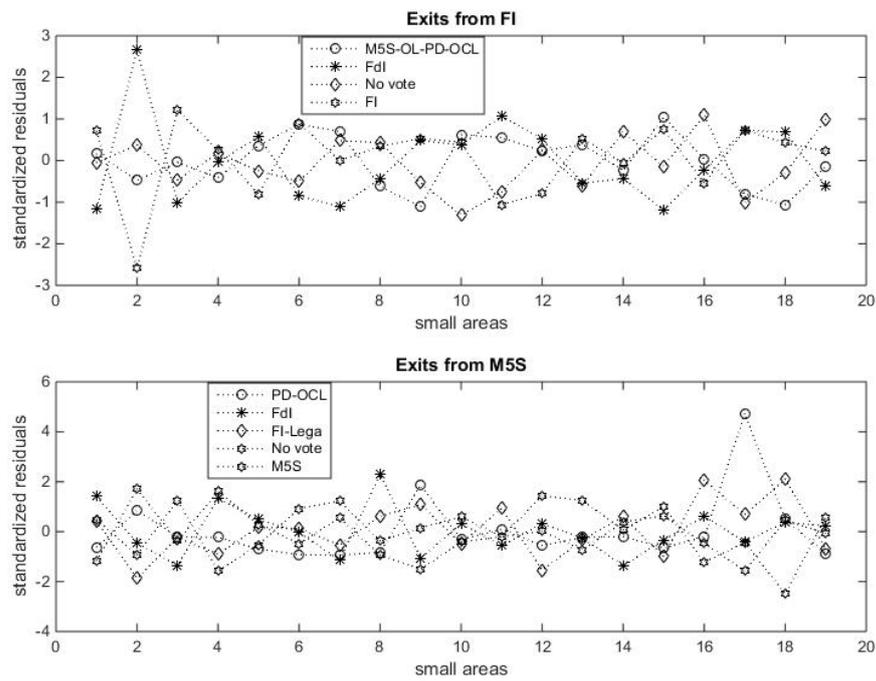

[a] *Party's acronyms: see Appendix B*